# Influence of intrinsic decoherence on tripartite entanglement and bipartite fidelity of polar molecules in pendular states

Jia-Xing Han(韩家兴)[1], Yuan Hu(胡远)[1], Yu Jin(金瑜)[1], Guo-Feng Zhang(张国锋) [1,2,3,4,*]

[1]*Key Laboratory of Micro-Nano Measurement-Manipulation and Physics (Ministry of Education), School of Physics and Nuclear Energy Engineering, Beihang University, Xueyuan Road No. 37, Beijing 100191, China*

[2]*State Key Laboratory of Software Development Environment, Beihang University, Xueyuan Road No. 37, Beijing 100191, China*

[3]*State Key Laboratory of Low-Dimensional Quantum Physics, Tsinghua University, Beijing100084, China*

[4]*Key Laboratory of Quantum Information, University of Science and Technology of China, Chinese Academy of Sciences, Hefei 230026, China*

An array of ultracold polar molecules trapped in an external electric field is regarded as a promising carrier of quantum information. Under the action of this field, molecules are compelled to undergo pendular oscillations by the Stark effect. Particular attention has been paid to the influence of intrinsic decoherence on the model of linear polar molecular pendular states, thereby we evaluate the tripartite entanglement with negativity, as well as fidelity of bipartite quantum systems for input and output signals using electric dipole moments of polar molecules as qubits. According to this study, we consider three typical initial states for both systems respectively, and investigate the temporal evolution with variable values of the external field intensity, the intrinsic decoherence factor and the dipole-dipole interaction. Thus we demonstrate the sound selection of these three main parameters to obtain the best entanglement degree and fidelity.

## I. INTRODUCTION

Since A. Einstein, B. Podolsky and N. Rosen raised the famous EPR paradox in 1935, followed by Schrödinger identifying this spooky action at a distance as the core of quantum mechanics, quantum entanglement has been attracting a critical mass of scientists to probe into the cryptic characteristics in quantum mechanics that are extremely different from the classical concept. A state of a composite systems is said to be entangled if its density matrix cannot be factored into a product of density matrices of the individual subsystems. Recently much effort has been devoted to understanding the dynamics of entanglement.[1-6] In quantum-information theory entanglement is a major resource for some practical applications and prominent theoretical efforts have been devoted both to the quantum algorithm designs and the physical realization.[7-14] Moreover a variety of potential qubit systems have being studied as a carrier of quantum information.[15-23] These include one-dimensional arrays of localized spins, coupled through exchange interactions and subject to an external magnetic field[21] and analogous treatments of trapped electric dipoles coupled by dipole-dipole interactions[22-23].

---

[*] Author to whom correspondence should be addressed.　Electronic mail: gf1978zhang@buaa.edu.cn.

Meanwhile, people have attained remarkable achievements in manufacturing, cooling and manipulating polar molecules, which combine the advantages of neutral atoms and trapped ions, considered as a promising new carrier in quantum information processing.[24-26] DeMille originally proposed a physical realization of quantum computer using ultracold polar molecules in 2002.[27] In this proposal, diatomic molecules are individually held in a 1D trap array, and the qubits consist of the electric dipole moments of these molecules, orienting along or against an external field, while they are entangled due to the dipole-dipole interaction. Polar molecules are in harmonic oscillation under the effect of external electric field, which is called the pendular state. As the vital resource of the application in the field of quantum information, however sometimes quantum interference and entanglement are extremely fragile and entangled states suffer severely from decoherence.[28-29] So it is crucial to preserve coherence to perform reversible quantum logic operations.[26] In DeMille's article,[27] several decoherence sources, including photon scattering from the trap laser, technical noise issue and other less nontrivial ones, have been briefly discussed with the result that they all appear controllable at the desired level and seem to present no limitations. In this paper, we specially consider the intrinsic decoherence effect proposed by Milburn, namely, the coherence decays automatically as the physical properties of a system approaching a macroscopic level.[30] Milburn decoherence gives a simple modification of unitary Schrödinger evolution which assumes that on sufficiently small time scales, the system evolves in a stochastic sequence of identical unitary transformations.

Multipartite pure-state is of great interest in the study of entanglement, on account that people tend to require a better understanding of the correlations between different registers of a quantum computer.[10] In this paper, we consider the model of three diatomic polar molecules in pendular state, and use Milburn's intrinsic decoherence model under the Markov's approximation, on which case we discuss the evolution of dynamic characteristics of the tripartite system. Here, we evaluate entanglement by negativity, an efficiently computable measure introduced by Vidal and Werner, applicative not only for bipartite mixed states but also multipartite entanglement.[31] Besides, we examine an aspect associated to but distinct from entanglement. The concept of fidelity is a basic ingredient in communication theory. For any given communication protocol, fidelity is a quantification of the similarity between the input and output quantum states.[32] With the parallel model, we calculate the variation of fidelity for bipartite quantum system over time, when certain parameters, such as intrinsic decoherence factor, dipole-dipole coupling interaction and external field strength, take different values.

## II. METHODS

### A. Tripartite negativity

G. Vidal and R. F. Werner constructed initially a useful quantity—negativity, which can be used to quantify the entanglement of an arbitrary bipartite system for any state.[31] Considering a bipartite pure state $|\psi_{AB}\rangle$, of which the density matrix is $\boldsymbol{\rho}$, the negativity for this system is defined as

$$N(\boldsymbol{\rho}) \equiv \sum_i |\lambda_i| - 1, \qquad (1)$$

which vanishes for any separable state, and where $\lambda_i$ are the eigenvalue of the partial transpose $\boldsymbol{\rho}^{T_A}$. By looking at the different bipartite splittings of a tripartite quantum system, we can classify effectively the entanglement properties of the system.[33-34] Hence negativity for the system can be written as the geometric average of that for any two component parts of this system[35]

$$N(\boldsymbol{\rho}_{ABC}) = \sqrt[3]{N_{AB-C} N_{AC-B} N_{BC-A}}, \qquad (2)$$

which reduces to the negativity of any two different component parts for a symmetric tripartite system.

### B. Fidelity of two-body system

For any given communication scheme, fidelity is a measure of the accuracy of transmission. In quantum information field, fidelity is used to quantify the similarity between input and output quantum states. Uhlmann introduced a notion of 'transition probability' for mixed states:[36]

$$F(\boldsymbol{\rho}_{in}, \boldsymbol{\rho}_{out}) = \left\{tr\left[\sqrt{(\boldsymbol{\rho}_{in})^{1/2}\boldsymbol{\rho}_{out}(\boldsymbol{\rho}_{in})^{1/2}}\right]\right\}^2 \quad (3)$$

Then Jozsa proposed this notion as a definition of fidelity for mixed quantum states and proved its essential properties. For instance, $0 \leq F(\boldsymbol{\rho}_{in}, \boldsymbol{\rho}_{out}) \leq 1$ and $F(\boldsymbol{\rho}_{in}, \boldsymbol{\rho}_{out}) = 1$ if and only if $\boldsymbol{\rho}_{in} = \boldsymbol{\rho}_{out}$.[32] On the contrary, $F(\boldsymbol{\rho}_{in}, \boldsymbol{\rho}_{out}) = 0$ if and only if $\boldsymbol{\rho}_{in}$ and $\boldsymbol{\rho}_{out}$ are completely orthogonal. The perfect fidelity under classical condition is 2/3, which indicates quantum information transmission will show its superiority when the fidelity is higher than 2/3. In terms of the communication scheme introduced by Jinhyoung Lee and M. S. Kim,[37] for given input quantum state $\boldsymbol{\rho}_{in}$, we will obtain the output quantum state $\boldsymbol{\rho}_{out}$ through the following formula,

$$\boldsymbol{\rho}_{out} = \sum_{i,j} p_{ij}(\boldsymbol{\sigma}_i \otimes \boldsymbol{\sigma}_j)\boldsymbol{\rho}_{in}(\boldsymbol{\sigma}_i \otimes \boldsymbol{\sigma}_j) \quad (4)$$

where $i, j = 0, 1, 2$ or $3$, $p_{ij} = tr[E^i\rho(t)]tr[E^j\rho(t)]$, $\boldsymbol{\rho}_{in} = |\psi\rangle_{in}\langle\psi|$, and

$$\begin{cases} |\psi^\pm\rangle = \frac{1}{\sqrt{2}}(|01\rangle \pm |10\rangle) \\ |\phi^\pm\rangle = \frac{1}{\sqrt{2}}(|00\rangle \pm |11\rangle) \end{cases}$$

$$\begin{cases} E^0 = |\psi^-\rangle\langle\psi^-| \\ E^1 = |\phi^-\rangle\langle\phi^-| \\ E^2 = |\phi^+\rangle\langle\phi^+| \\ E^3 = |\psi^+\rangle\langle\psi^+| \end{cases}$$

## C. The model of intrinsic decoherence

Milburn has proposed a simple model of intrinsic decoherence by giving a modification of standard quantum mechanics. He assumed that on sufficiently short time steps the system does not evolves continuously under unitary evolution but rather in a stochastic sequence of identical unitary phase changes.[30] In other words, on a sufficiently small time scale $\tau$, the state of the system changes by

$$\boldsymbol{\rho}(t + \tau) = \exp\left[-\frac{i}{\hbar}\theta(\tau)H\right]\rho(t)\exp\left[\frac{i}{\hbar}\theta(\tau)H\right], \quad (5)$$

with a probability of p(t). In the following, we will take $\hbar = 1$. There is some minimum unitary phase change because of the setting $\lim_{\tau\to 0}\theta(\tau) = \lim_{\tau\to 0}\tau/p(t) = \theta_0$. Effectively this introduces a minimum time step in the universe. The coherence is automatically destroyed as the physical properties of the system approach a macroscopic level. So this might be called "intrinsic decoherence". Based on these assumptions, Milburn derived an equation for the time evolution density operator ρ(t) of the quantum system. Then, by expanding this equation to first order in the expansion parameter, he obtained the following Milburn equation:

$$\frac{d\boldsymbol{\rho}(t)}{dt} = -i[\boldsymbol{H}, \boldsymbol{\rho}(t)] - \frac{\gamma}{2}[\boldsymbol{H}, [\boldsymbol{H}, \boldsymbol{\rho}(t)]], \qquad (6)$$

where $\gamma$ is the intrinsic decoherence factor and corresponds to the $\gamma^{-1}$ in Milburn's paper. Thus in the limit that $\gamma \to 0$, the system evolves continuously and there is no intrinsic decoherence.

The model discussed above is intended to apply both to individual systems prepared initially in pure states or to large collections of systems perhaps better described initially as mixed states. Milburn has investigated the Milburn equation for various systems and proved its applicability.[30] However, note that this equation will only be true if the identical unitary transformations are maintained as unitary phase changes and that is the applicable physical channel to our intrinsic decoherence model. Xu obtained the formal solution of the Milburn equation as follows,[29]

$$\boldsymbol{\rho}(t) = \sum_{k=0}^{\infty} \frac{(\gamma t)^k}{k!} \boldsymbol{M}^k \boldsymbol{\rho}(0) \boldsymbol{M}^{\dagger k}, \qquad (7)$$

where $\boldsymbol{\rho}(0)$ is the density operator of the initial system and $\boldsymbol{M}^k$ has been defined by

$$\boldsymbol{M}^k = \boldsymbol{H}^k e^{-i\boldsymbol{H}t} e^{-\frac{\gamma t}{2}\boldsymbol{H}^2} \qquad (8)$$

Thus, for the time evolution density operator of the quantum system described by the Milburn equation can be written as

$$\boldsymbol{\rho}(t) = \sum_{m,n} \exp\left[-\frac{t\gamma}{2}(\varepsilon_m - \varepsilon_n)^2 - i(\varepsilon_m - \varepsilon_n)t\right] \times \langle \xi_m|\boldsymbol{\rho}(0)|\xi_n\rangle |\xi_m\rangle\langle\xi_n|, \qquad (9)$$

where $\varepsilon_{m,n}$ and $|\xi_{m,n}\rangle$ are the eigenvalues and corresponding eigenstates of the Hamiltonian of the system, respectively.

### III. THE MODEL

B. Friedrich and D.R. Herschbach[38] established a model for a rigid linear polar molecule which is trapped in an external electric field ε and obtained the Hamiltonian

$$H = \frac{p^2}{2m} + V_{trap}(r) + BJ^2 - \mu \cdot \varepsilon,$$

Where m, B and μ represent the mass, rotational constant and body-fixed dipole moment of the molecule respectively. $P^2/2m$ is translational kinetic energy and $V_{trap}(r)$ is potential energy of the molecule within the trapping field. When the molecule exists in an ultra-cold environment, its translational motion will be quite modest and nearly harmonic. $P^2/2m + V_{trap}(r)$ thus can be omitted from the Hamiltonian as a constant. There will remain the rotational kinetic energy and the Stark interaction term which appeared in our manuscript as

$$H_s = BJ^2 - \mu\varepsilon\cos\theta,$$

where θ is the polar angle between the molecular axis and the additional electric field direction. Both the eigenvalues and the corresponding eigenfunctions of $\boldsymbol{H_s}$ have respectively been obtained by von Meyenn and Stejskal.[39-40] Since the angular momentum $\boldsymbol{J}$ and the Hamiltonian are noncommutative, $J$ is no longer a good quantum number. However, $M$, the projection of the angular momentum vector $\boldsymbol{J}$ on the $z$ axis, retains azimuthal symmetry about the direction of additional electric field. We can still consider $M$ a good quantum number and identify the eigenstates with $M$ and $J$. D. DeMille described the molecular qubits as elctric dipole moments oriented along ($|0\rangle$) or against ($|1\rangle$) an external electric field, which correspond the $M=0$, $J=0$ and $M=0$, $J=1$ for the mixed state with low fields, respectively.[27] The qubits can be formulated as linear superposition of spherical harmonics,[22]

$$|0\rangle = \sum_i a_i Y_{i,0}(\theta,\varphi), \quad |1\rangle = \sum_j a_j Y_{j,0}(\theta,\varphi).$$

**A. Two dipoles in pendular states**

Adding a second trapped polar molecule, the Hamiltonian is identical to the first except the interaction energy. C. Ticknor and J. L. Bohn formulated the dipole-dipole interaction in vector form as follows,

$$H_{1,2} = \frac{\boldsymbol{\mu_1}\cdot\boldsymbol{\mu_2} - 3(\boldsymbol{\mu_1}\cdot\boldsymbol{n})(\boldsymbol{\mu_2}\cdot\boldsymbol{n})}{|r_1-r_2|^3}, \quad (10)$$

where $\boldsymbol{n}$ stands for the unit vector of $\boldsymbol{r}_{12}$, $\boldsymbol{\mu}_1$ and $\boldsymbol{\mu}_2$ stand for the electric dipole moment of polar molecules.[41] For $M=0$ states whose dipole-dipole interaction is angular symmetry, the dipole-dipole interaction can be simplified by averaging over the angle:

$$V_{1,2} = \Omega(1 - 3\cos^2\alpha)\cos\theta_1\cos\theta_2, \quad (11)$$

where $\Omega = \boldsymbol{\mu}^2/r_{1,2}^3$, $\alpha$ stands for the angle between vector $\boldsymbol{r}_{12}$ and the electric field direction, $\theta_1$ and $\theta_2$ are the angle of electric dipole moment $\boldsymbol{\mu}_1$, $\boldsymbol{\mu}_2$ and the electric field direction. The Hamiltonian

$$H_s = H_{s1} + H_{s2} + H_{1,2}$$

when set up in a basis of the qubits, $\{|00\rangle, |01\rangle, |10\rangle, |11\rangle\}$, take the form

$$H_{s1} = \begin{pmatrix} E_0^1 & 0 \\ 0 & E_1^1 \end{pmatrix} \otimes I_2,$$

$$H_{s2} = I_1 \otimes \begin{pmatrix} E_0^2 & 0 \\ 0 & E_1^2 \end{pmatrix},$$

$$H_{1,2} = \Omega_{12} \left[ \begin{pmatrix} C_0^1 & C_t^1 \\ C_t^1 & C_1^1 \end{pmatrix} \otimes \begin{pmatrix} C_0^2 & C_t^2 \\ C_t^2 & C_1^2 \end{pmatrix} \right],$$

where $E_0^x$ and $E_1^x$ ($x$=1, 2) are the eigenenergies of qubits $|0\rangle$ and $|1\rangle$, $I_x$ is a unit matrix of $2 \times 2$ dimension. The matrix elements of $H_{1,2}$ can be formulated as

$$C_0^x = \langle 0|\cos\theta|0\rangle,$$
$$C_t^x = \langle 0|\cos\theta|1\rangle,$$
$$C_1^x = \langle 1|\cos\theta|1\rangle,$$

where $C_0^x$ and $C_1^x$ are the expected values of $\cos\theta$ in the qubits $|0\rangle$ and $|1\rangle$, respectively, which are the orientations of molecules. $C_t^x$ corresponds to an exchange interaction or transition dipole moment between the qubit states.

**B. Three dipoles in pendular states**

For the Hamiltonian of three polar molecules trapped by the additional electric field, it will become more complex. This paper assumes three identical molecules positioned in a linear and isometric array, so the dipole–dipole coupling interaction of arbitrary two polar molecules and the Hamiltonian of the system are analogous to the situation for two polar molecules. The Hamiltonian

$$H_s = H_{s1} + H_{s2} + H_{s3} + H_{1,2} + H_{1,3} + H_{2,3},$$

when set up in a basis of the qubits, $\{|000\rangle, |001\rangle, |010\rangle, ... |111\rangle\}$, take the form

$$H_{s1} = \begin{pmatrix} E_0^1 & 0 \\ 0 & E_1^1 \end{pmatrix} \otimes I_2 \otimes I_3,$$

$$H_{s2} = I_1 \otimes \begin{pmatrix} E_0^2 & 0 \\ 0 & E_1^2 \end{pmatrix} \otimes I_3,$$

$$H_{s3} = I_1 \otimes I_2 \otimes \begin{pmatrix} E_0^3 & 0 \\ 0 & E_1^3 \end{pmatrix},$$

$$V_{12} = \Omega_{12} \left[ \begin{pmatrix} C_0^1 & C_t^1 \\ C_t^1 & C_1^1 \end{pmatrix} \otimes \begin{pmatrix} C_0^2 & C_t^2 \\ C_t^2 & C_1^2 \end{pmatrix} \otimes I_3 \right],$$

$$V_{13} = \Omega_{13}\left[\begin{pmatrix} C_0^1 & C_t^1 \\ C_t^1 & C_1^1 \end{pmatrix} \otimes I_2 \otimes \begin{pmatrix} C_0^3 & C_t^3 \\ C_t^3 & C_1^3 \end{pmatrix}\right],$$

$$V_{23} = \Omega_{22}\left[I_1 \otimes \begin{pmatrix} C_0^2 & C_t^2 \\ C_t^2 & C_1^2 \end{pmatrix} \otimes \begin{pmatrix} C_0^3 & C_t^3 \\ C_t^3 & C_1^3 \end{pmatrix}\right].$$

## IV. RESULTS

### A. Tripartite negativities of polar molecules in pendular states

We consider three different initial states for the tripartite system: (a) GHZ state $|\Psi\rangle = \frac{1}{\sqrt{2}}(|000\rangle + |111\rangle)$, (b) W state $|\Psi\rangle = \frac{1}{\sqrt{3}}(|001\rangle + |010\rangle + |100\rangle)$ and (c) separable state $|\Psi\rangle = |001\rangle$ that will evolve under the influence of intrinsic decoherence respectively. Then we will study how these negativities change over time $t$ with $\gamma$ the factor of intrinsic decoherence, $\varepsilon$ the additional electric field strength and $\Omega$ the dipole–dipole coupling interaction variable. This paper assumes three identical molecules positioned in a linear and isometric array, that is $r_{1,3} = 2r_{1,2} = 2r_{1,2}$. $\Omega_{12} = \Omega_{23} = 8\Omega_{13}$ since $\Omega = \mu^2/r^3$ which is defined in eq.10. In the following numerical calculation, we have assumed $\Omega = \Omega_{12} = \Omega_{23} = 8\Omega_{13}$ and that $\varepsilon$ is perpendicular to the linear chain vector $r_{xy}$ of the three polar molecules, namely, the angle $\alpha$ between these two vectors is $\pi/2$.

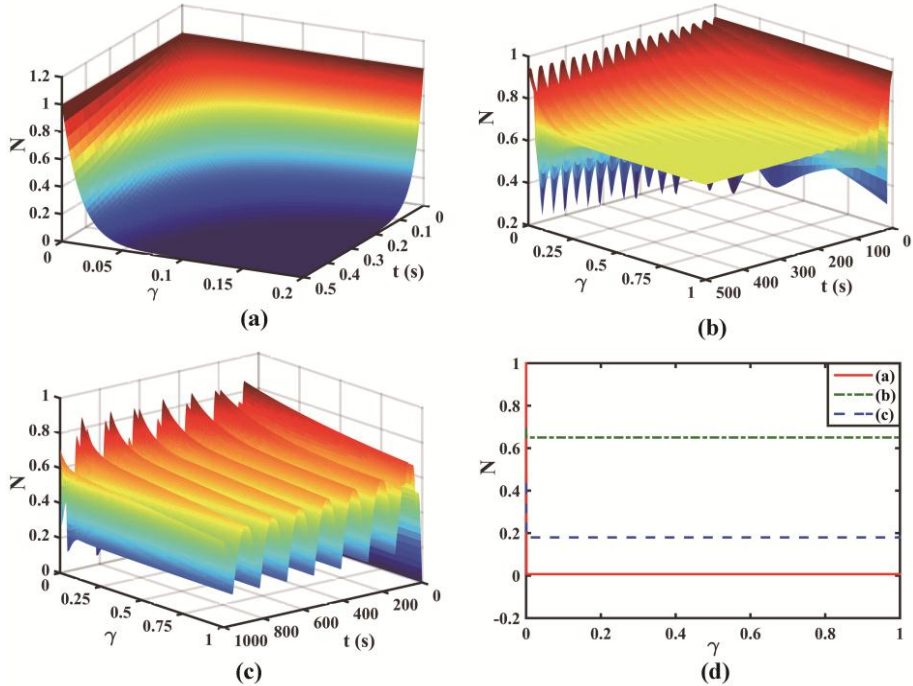

FIG. 1. Negativities of tripartite systems as functions of $\gamma$ and time $t$ with $\mu\varepsilon/B = 6$, $\Omega/B = 0.5$ for different initial states: (a) GHZ state $|\Psi\rangle = (|000\rangle + |111\rangle)/\sqrt{2}$; (b) W

state $|\Psi\rangle = (|001\rangle + |010\rangle + |100\rangle)/\sqrt{3}$ and (c) separable state $|\Psi\rangle = |001\rangle$. Negativities for values of $\gamma$ when $t$ becomes long enough are displayed in (d).

As plotted in Figure 1 (a), negativities remain 1 when the decoherence is completely suppressed ($\gamma = 0$) since the negativities of initial GHZ state is maximally entangled. However, negativities drop rapidly to be zero in a remarkably short period of time as $\gamma$ is introduced, and there is a positive correlation between the decrease and $\gamma$, because $\gamma$ represents the coupling strength between the tripartite system and the external environment which causes the intrinsic decoherence.

Figure 1 (b), (c) and (d) show that when $t = 0$, negativities of the system (b) and (c) are respectively 0.94 and 0 for the reason that initial W state is partially entangled and separable state is non-entangled. When $t > 0$, negativities vibrate with a $\gamma$-independent period and tend respectively to a constant that is independent of $\gamma$ for a sufficiently long time as $\gamma > 0$. Besides, the speeds of vibration attenuation grow with $\gamma$ and it develops a continuous oscillation as $\gamma = 0$ for the similar reasons as for GHZ state. In terms of the similarity among (a), (b) and (c), we naturally believe that when $\gamma = 0$, the change of negativities over time for GHZ state could be regarded as a special case of continuous oscillation in which the amplitude equals to zero. Some of the change rules mentioned above can also be drawn from the evolution formula $\boldsymbol{\rho}(t) = \sum_{m,n} \exp\left[-\frac{t}{2\gamma}(\varepsilon_m - \varepsilon_n)^2 - i(\varepsilon_m - \varepsilon_n)t\right] \times \langle\zeta_m|\boldsymbol{\rho}(0)|\zeta_n\rangle|\zeta_m\rangle\langle\zeta_n|$ qualitatively, where $\exp\left[-\frac{t}{2\gamma}(\varepsilon_m - \varepsilon_n)^2 - i(\varepsilon_m - \varepsilon_n)t\right]$ is related to the vibration of which the frequency is $(\varepsilon_m - \varepsilon_n)$ independent of $\gamma$. When $\gamma = 0$, $\exp[-i(\varepsilon_m - \varepsilon_n)t]$ represents an equal amplitude oscillation. When $\gamma > 0$, $\exp\left[-\frac{t}{2\gamma}(\varepsilon_m - \varepsilon_n)^2\right]$ represents the attenuation of amplitude and the speed of attenuation grows with $\gamma$. Especially the amplitude tends to zero as $t \to \infty$.

What we have discussed above reveals the fact that the intrinsic decoherence just affects the speed of the evolution but not the final state of the system. However, the choice of initial state does have a great effect on the final state. Apparently, as displayed is figure (d), the constant of negativities for initial state (b) is more than three times for initial state (c), while it is almost zero for initial state (c). This shows the advantage of partially entangled state—W state.

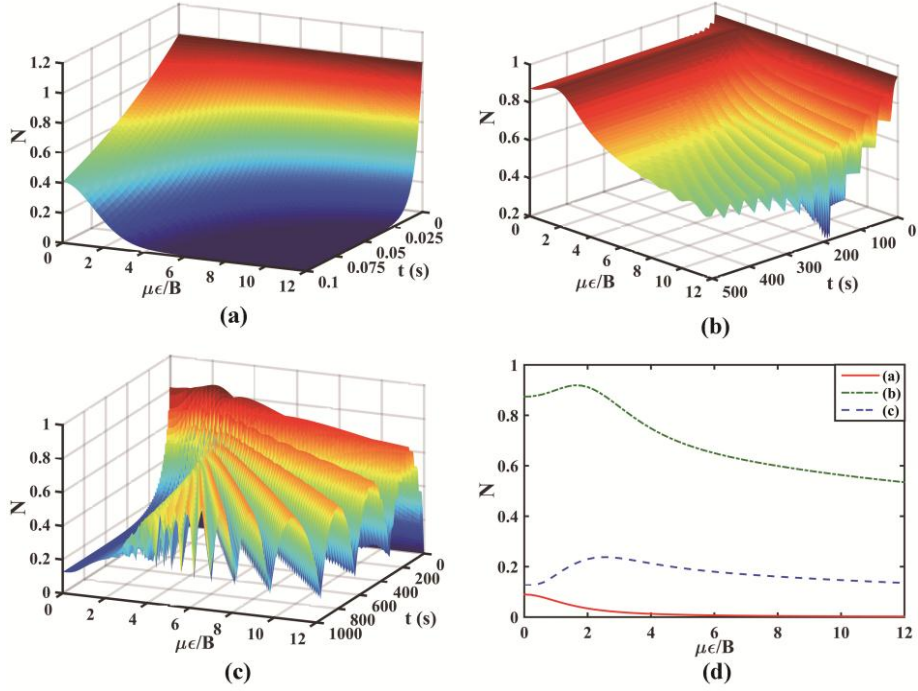

FIG. 2. Negativities of tripartite systems as functions of $\mu\varepsilon/B$ and time $t$ with $\gamma = 0.5, \Omega/B = 0.5$, for differnent initial states: (a) GHZ state $|\Psi\rangle = (|000\rangle + |111\rangle)/\sqrt{2}$; (b) W state $|\Psi\rangle = (|001\rangle + |010\rangle + |100\rangle)/\sqrt{3}$ and (c) Separable state $|\Psi\rangle = |001\rangle$. Negativities for values of $\mu\varepsilon/B$ when $t$ becomes long enough are displayed in (d).

As can be seen in figure 2 (a), the change trend of negativities appears similar to figure 1(a) except that negativities decline over time for $\mu\varepsilon/B = 0$ case because of the existence of dipole-dipole interaction. Qualitatively, the more powerful interaction occurs between the external field and the system, the faster the decoherence of the system is. Figure (b) and (c) plot that negativities vary with time (the intensity is generally positively correlated with $\mu\varepsilon/B$) and tend to a steady value which depends on the value of $\mu\varepsilon/B$. These dependencies are, more clearly, shown in figure (d) where initial W state shows the advantage again. For initial state (a), negativities are virtually nil as $\mu\varepsilon/B > 5$, but for initial state (b), negativities have a maximum when $\mu\varepsilon/B$ is about 1.6 and always keep a much higher value than that for initial state (c), of which the curve seems to have a similar shape. For initial W state, 1.6 is the perfect value of $\mu\varepsilon/B$ because of its highest stable negativity and shortest time needed for the negativity to tend to be a constant.

According to perturbation theory of nondegenerate states, we can roughly analyse the effect of external field. In the first order approximation, eigenvalues and eigenvectors of Hamiltonian for a polar molecule are $\varepsilon_0 = A\left[2 - \frac{2}{3}\left(\frac{\mu\varepsilon}{B}\right)^2\right], \varepsilon_1 = A, |0\rangle = |Y_{00}\rangle + \sqrt{\frac{1}{3}\frac{\mu\varepsilon}{B}}|Y_{10}\rangle, |1\rangle = |Y_{10}\rangle - \sqrt{\frac{1}{3}\frac{\mu\varepsilon}{B}}|Y_{00}\rangle$, where A is a function of $\mu\varepsilon/B$. When $\varepsilon_0 = \varepsilon_1$, namely $\mu\varepsilon/B = 1.22$, $\boldsymbol{\rho}(t)$ does not change over time and takes a maximum value.

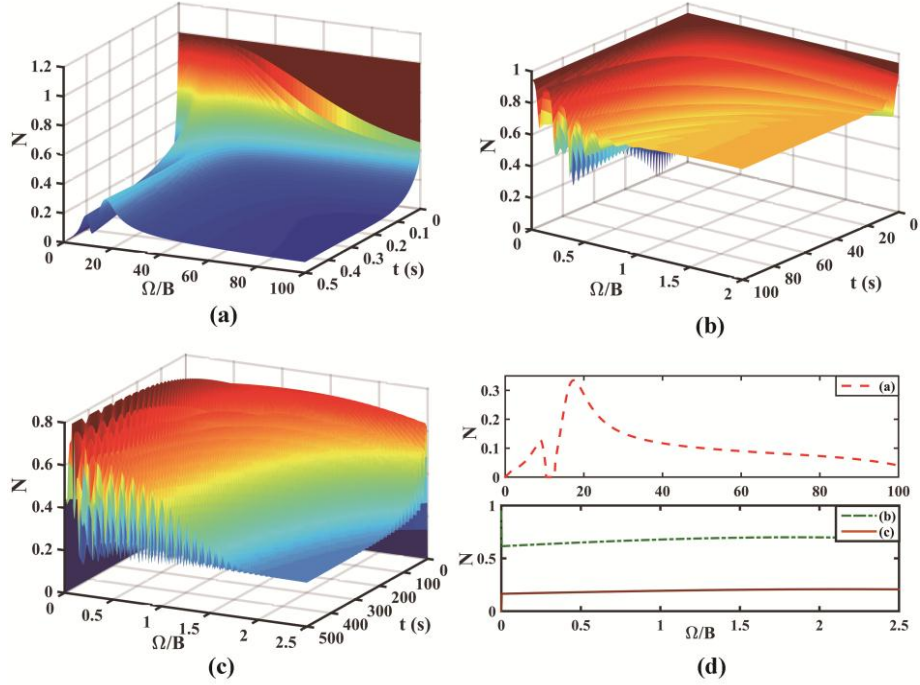

FIG. 3. Negativities of tripartite systems as functions of $\Omega/B$ and time $t$ with $\gamma = 0.5$, $\mu\varepsilon/B = 6$ for different initial states: (a) GHZ state $|\Psi\rangle = (|000\rangle + |111\rangle)/\sqrt{2}$; (b) W state $|\Psi\rangle = (|001\rangle + |010\rangle + |100\rangle)/\sqrt{3}$ and (c) Separable state $|\Psi\rangle = |001\rangle$. Negativities for values of $\Omega/B$ when $t$ becomes long enough are displayed in (d).

For initial state (a) in Figure 3, negativities drop rapidly but there are two stable peaks at $\Omega/B = 9$, 17.5 respectively. For initial state (b) and (c), negativities vary with time and tend to a steady value which is dependent on the value of $\Omega/B$. Note that the frequency of the vibration is exactly proportional to $\Omega/B$ because $\Omega/B$ is proportional to $(\varepsilon_m - \varepsilon_n)$ according to the structure of the Hamiltonian matrix for the system. Moreover, negativities for initial state (b) are, overall, higher than 0.6 and much better than that for initial state (a) and (c).

## B. Bipartite fidelities of polar molecules in pendular states

We assume that the initial state of the two-body system is $|\Psi\rangle = a|01\rangle + b|10\rangle$, which will evolve under the influence of intrinsic decoherence, where $a$ and $b$ are arbitrary constants and satisfy the normalization condition. We take $\boldsymbol{\rho}(t)$ the density matrix of the system for every moment $t$ as the input states of the communication scheme as noted above and investigate how the fidelities change over time $t$ with $\gamma$ the factor of intrinsic decoherence, $\varepsilon$ the additional electric field strength and $\Omega$ the dipole–dipole coupling interaction variable.

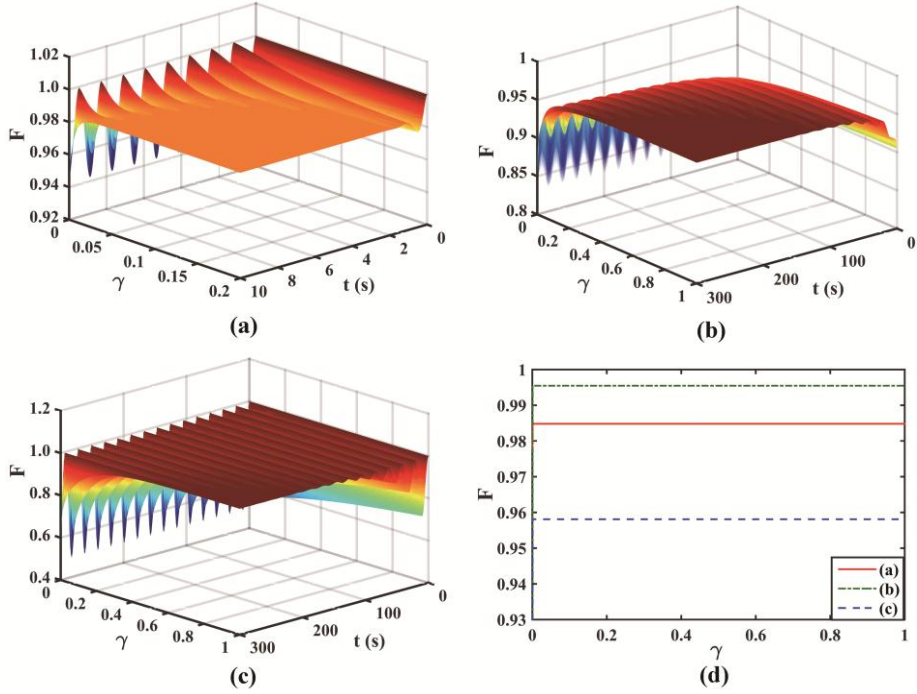

FIG. 4. Fidelities of bipartite systems as functions of the intrinsic decoherence factor $\gamma$ and time $t$ with $\mu\varepsilon/B = 6$, $\Omega/B = 1$ for differnent initial states: (a) $a = b = \sqrt{2}/2$; (b) $a = \sqrt{3}/2, b = 1/2$ and (c) $a = 1, b = 0$. Fidelities for values of $\gamma$ when $t$ becomes long enough are displayed in (d).

As can be seen in figure 4, trends in the evolution of fidelities for the three different initial states seem similar in some respects. Note that fidelities for each initial state vibrate periodically over time and incline to a constant, where the periods and the constants are independent of $\gamma$. We can quantitatively analyse these regularities by the same way as used in figure 1. And meanwhile, the stable value of fidelities for initial state (b) is marginally higher than that for the others and almost close to 1. However, the speeds of vibration attenuation vary among different initial states (for (a), the attenuation speed is almost 1000 times faster than that for the others) and grow with $\gamma$, for which it will develop a continuous oscillation as $\gamma = 0$. Overall, the intrinsic decoherence just affects the speed of the evolution of but not the final value of fidelity, and maximally entangled state (a) shows its advantage here for its shortest time needed for the fidelity to tend to be a constant.

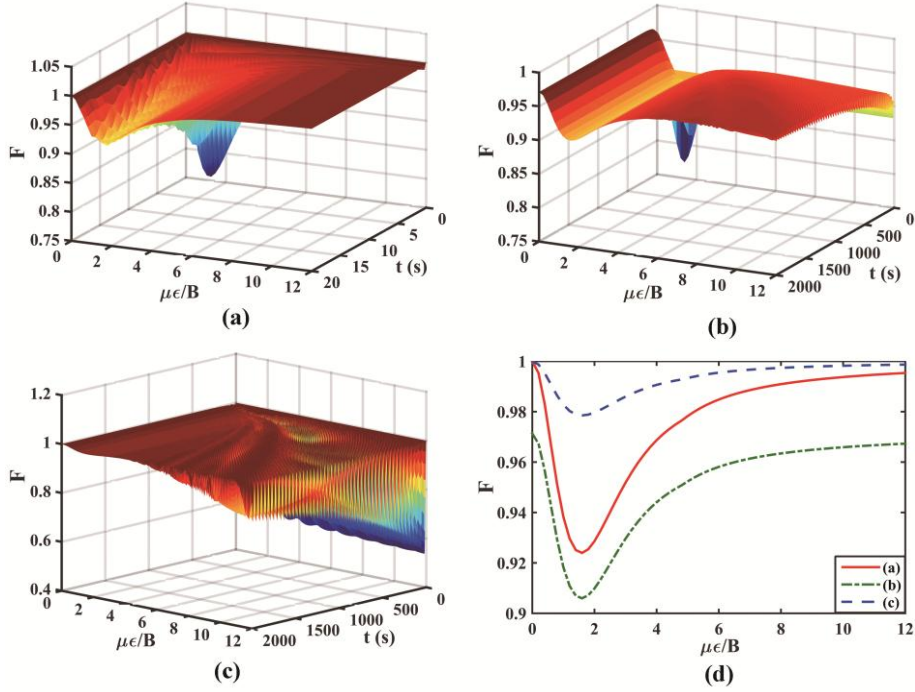

FIG. 5. Fidelities of bipartite systems as functions of $\mu\varepsilon/B = 6$ and time t with $\Omega/B = 1$, $\gamma = 0.05$ for different initial states: (a) $a = b = \sqrt{2}/2$; (b) $a = \sqrt{3}/2, b = 1/2$ and (c) $a = 1, b = 0$. Fidelities for values of $\mu\varepsilon/B$ when t becomes long enough are displayed in (d).

Figure 5 describes that fidelities of bipartite systems fluctuate over time and tend, for a sufficiently long time, to definite values which are dependent on different initial states and different values of $\mu\varepsilon/B$. As indicated in figure (d), fidelities steeply fall to a minimum when $\mu\varepsilon/B$ is about 1.6 and then rise steadily for all three initial states. Separable state (c) performs better here, and note that fidelities remain 1 for the separable initial state (c) and the maximally entangled initial state (a) without electric field.

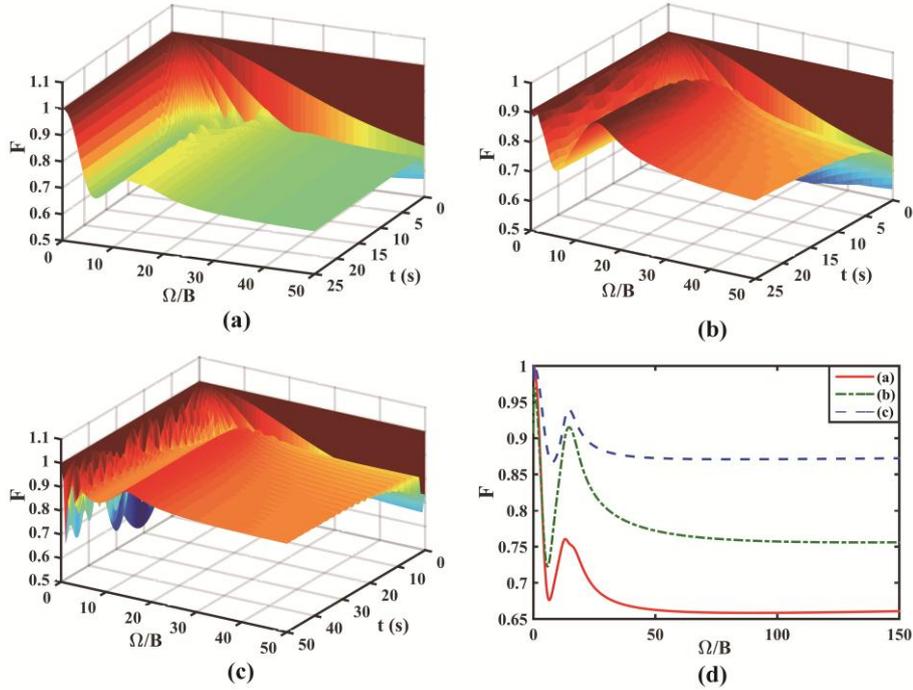

FIG. 6. Fidelities of bipartite systems as functions of $\Omega/B$ and time $t$ with $\mu\varepsilon/B = 6, \gamma = 0.05$, for differnent initial states: (a) $a = b = \sqrt{2}/2$; (b) $a = \sqrt{3}/2, b = 1/2$ and (c) $a = 1, b = 0$. Fidelities for values of $\Omega/B$ when t becomes long enough are displayed in (d).

As presented in figure 6 fidelities fluctuate over time and tend to a definite value, which is dependent on different initial states and different values of $\Omega/B$. Figure (d) shows that the stable values of fidelities tend to decline as a whole with except for a kink with a local maximum when $\Omega/B$ is between 10 and 20 for each initial state respectively. And note that the fidelities remain 1 without dipole-dipole coupling interaction ($\Omega/B = 0$) for all three initial states.

## V. CONCLUSIONS

In this paper, our chief aim has been to examine entanglement of polar molecules in linear equidistant distribution, which are subject to an external electric field and coupled to each other. Previous studies have explicitly presented the entanglement of systems consisting of two polar molecules. However, we should envisage the problem of multiple entanglement that represents the next order of complexity in this field. Meanwhile, when physical realization is concerned for arbitrary quantum system, the influence of intrinsic decoherence ought to be taken into account, which rarely occurred in the research for quantum application of ultracold molecules antecedently. We evaluate how the destruction of coherence infects the entanglement of three coupling polar molecules since the factor of intrinsic decoherence are introduced to modify the original system.

We calculated the tripartite negativity as a function of three simplified variables involving the dipole moment, the field strength, the rotational constant, the dipole-dipole interaction and the intrinsic decoherence factor. By manipulating these parameters, we observed how the negativities evolved with time and made several useful conclusions. Although GHZ state is a maximal entangled state, its properties under the influence of intrinsic decoherence are, in general, bad for its rapidly complete disentanglement. Fortunately, W state's negativities exhibit a good initial value and a much higher stable value, so we can chose W state as initial state to obtain a powerful entanglement of the system. Meanwhile, it is almost certain that the largest impact of the stable values of negativities is $\mu\varepsilon/B$, which can be used to reach the maximal entanglement.

In the parallel aspect, we have also studied fidelities for the given communication scheme, as previously mentioned. We can plainly derive from the limited cases analyzed above, that fidelities are almost always higher than **2/3**—the perfect value under the classical circumstance. Namely, it satisfies fundamental requirement for quantum information transmission. Furthermore, by choosing suitable parameters $\gamma$, $\varepsilon$ and $\Omega$ and designating an appropriate initial state, the value of fidelities can be even more excellent, simultaneously shorten the stabilizing time of the systems. For example, choosing initial state (a) will effectively reduce the stabilizing time of the system and we should avoid setting $\mu\varepsilon/B$ the value around 1.6. However, transmission of quantum information will lose its advantage for the maximally entangled initial state (a) when $\Omega/B > 50$, which might be an unexpected result.

How to put the research into practice, however, faces many obstacles ahead. We seem to have no access of the genuine value of the intrinsic decoherence factor, not to mention the manipulation of it. In the case that $\gamma$ levels off to $10^{-1}$, we can observe clearly that the negativities and fidelities steeply approach to zero, which directly restricts the application on quantum teleportation. Fortunately, on the other hand, quantum computation does not require an intense entanglement. We can still prepare multibit phase gates using ultracold polar molecules. Methods of enhancing the dipole-dipole interaction also remain to be researched.

Both the classification of entanglement and its quantification are complicated in mathematical computation. So we restricted ourselves up to the three polar molecules and most of the results in the previous section are for these systems. Although multipartite entanglement in many-body systems is much less studied, and the structure of entanglement in many-body systems is much richer, it is natural to expect that the results of several quantities like negativity will be useful for multipartite entanglement when the whole system has a symmetry, since it can be abtained based on the factorizability of a given many particle state into smaller parts. For more detail with sufficient evidence, further work is thus required.

## ACKNOWLEDGEMENT

This work is supported by the Student Innovation and Pioneering Training Program and the National Natural Science Foundation of China (Grant No. 11574022 and 11174024).